\documentclass[a4paper,12pt]{article}
\usepackage[super,square]{natbib}
\usepackage{anysize}
\marginsize{2cm}{2cm}{1cm}{2cm}
\usepackage{fancyhdr}

\usepackage{soul}
\usepackage[dvipsnames]{xcolor}
\usepackage{graphicx}
\usepackage{caption}
\usepackage{amsmath}
\usepackage{hyperref}
\hypersetup{
    colorlinks=true,
    linkcolor=black,
    citecolor=CadetBlue,
    filecolor=CadetBlue,      
    urlcolor=CadetBlue,
}
\usepackage{multicol}
\usepackage{parskip}
\renewenvironment{abstract}
 {\par\noindent\textbf{\abstractname}\ \ignorespaces \\}
 {\par\noindent\medskip}

\begin{document}
\pagestyle{fancy}
\thispagestyle{empty}
\fancyhead[L]{}
\begin{center}
\Large{\textbf{Predicting Flow Dynamics using Diffusion Models}}
\vspace{0.4cm}
\normalsize
\\ Yannick Gachnang, Vismay Churiwala \\
\vspace{0.1cm}
\textit{University of Pennsylvania}
\medskip
\normalsize
\end{center}
{\color{gray}\hrule}
\vspace{0.4cm}
\begin{abstract}
In this work, we aimed to replicate and extend the results presented in the DiffFluid paper\cite{luo2024difffluidplaindiffusionmodels}. The DiffFluid model showed that diffusion models combined with Transformers are capable of predicting fluid dynamics. It uses a denoising diffusion probabilistic model (DDPM) framework to tackle Navier-Stokes and Darcy flow equations. Our goal was to validate the reproducibility of the methods in the DiffFluid paper while testing its viability for other simulation types, particularly the Lattice Boltzmann method.

Despite our computational limitations and time constraints, this work provides evidence of the flexibility and potential of the model as a general-purpose solver for fluid dynamics. Our results show both the potential and challenges of applying diffusion models to complex fluid dynamics problems. This work highlights the opportunities for future research in optimizing the computational efficiency and scaling such models in broader domains.
\end{abstract}
{\color{gray}\hrule}
\medskip
\begin{multicols}{2}
\tableofcontents
\section{Introduction}

Efficient and accurate simulations of fluid dynamics are important in computational science and engineering, with applications in weather forecasting, aerospace design and more. While traditional numerical methods like finite difference, finite volume, and spectral techniques have delivered impressive precision, they often require significant computational resources and struggle to scale effectively. More recently however, machine learning has become a big area of interest for physics based simulations. Techniques such as neural operators and diffusion-based models have shown potential for providing fast, flexible, and accurate approximations of complex fluid behaviors.

In this work, we aimed to replicate and extend the results presented in the DiffFluid paper \cite{luo2024difffluidplaindiffusionmodels}. The DiffFluid model showed that diffusion models combined with Transformers are capable of predicting fluid dynamics. It uses a denoising diffusion probabilistic model (DDPM) framework to tackle Navier-Stokes and Darcy flow equations. Our goal was to validate the reproducibility of the methods in the DiffFluid paper while testing its viability for other simulation types, particularly the Lattice Boltzmann method.

\subsection{Motivation}

Machine learning-based fluid solvers hold great potential but face significant challenges. Models like Fourier Neural Operators (FNO) and Physics-Informed Neural Networks (PINNs) have achieved milestones but often struggle to generalize across diverse physical systems. Additionally, they can fail to capture the sharp, multiscale features characteristic of turbulent flows. DiffFluid offers a new perspective by treating flow prediction as a generative task of frame translation. It leverages multiscale noise injection and a Transformer backbone to address these challenges. 

Our replication study aimed to validate the robustness and reproducibility of the DiffFluid framework, ensuring that its performance could be reliably replicated. We also sought to investigate its application across diverse simulation methodologies to test its adaptability. Finally, we assessed how adjustments, including noise strategies and loss functions, influenced the overall performance of the framework.

\subsection{Simulation Methods}

For the Navier-Stokes simulations, we used an implementation based on the Fourier Neural Operator (FNO) paper\cite{li2021fourierneuraloperatorparametric}. This approach uses the Crank-Nicholson integration to solve the incompressible Navier-Stokes equations in vorticity form. 

In contrast, the Lattice Boltzmann method (LBM) offers an alternative approach with which we can verify whether the model can be adapted to other problems as well. LBM operates at a mesoscopic level, simulating particle distributions to recover macroscopic flow dynamics. We employed the D2Q9 lattice model, a common configuration for 2D flows, to simulate shear layer instability. Initial conditions included randomized velocity fields, with parameters tuned heuristically to ensure realistic flow evolution. This simulation method was easy to adapt, thus it made for a great testing candidate.

\subsection{Contributions}

Our study focused on validating the methodologies proposed in the DiffFluid paper and exploring their application in more simulation techniques. We reproduced DiffFluid's results on Navier-Stokes simulations and achieved comparable levels of visual quality. This confirms the robustness of the DiffFluid model. We also tested the adaptability of DiffFluid's diffusion-based model by applying it to Lattice Boltzmann simulations, which shows its capacity to handle alternative fluid dynamics simulations. And we explored ways to reduce computational demands by scaling down the dataset size and resolution without a large loss in information.

\end{multicols}
{\color{gray}\hrule}
\begin{center}
\section{Material \& Methods}
\bigskip
\end{center}
{\color{gray}\hrule}
\begin{multicols}{2}

\subsection{Theory}

The Navier-Stokes equations form the cornerstone of fluid dynamics, describing the motion of incompressible and viscous fluids. Represented in vorticity form, the equations are given by:

\[
\frac{\partial \omega}{\partial t} + (\mathbf{u} \cdot \nabla)\omega = \nu \nabla^2 \omega + f
\]

Here, \(\omega\) is the vorticity, \(\mathbf{u}\) is the velocity field derived from the vorticity, \(\nu\) is the kinematic viscosity, and \(f\) represents external forces. This inherently satisfies the incompressibility condition of liquids, \(\nabla \cdot \mathbf{u} = 0\). Solving these equations computationally is challenging due to the nonlinear advective term and the high Reynolds numbers encountered in turbulent flows.

The Lattice Boltzmann Method (LBM) is a numerical approach for simulating fluid flows. It is good at capturing mesoscopic-scale behaviors and handling complex boundaries. The standard D2Q9 lattice model was used in this study. The evolution of the particle distribution function, \( f_i \), was governed by the lattice Boltzmann equation:

\[
f_i(\mathbf{x} + \mathbf{e}_i \Delta t, t + \Delta t) = f_i(\mathbf{x}, t) - \frac{f_i(\mathbf{x}, t) - f_i^{\text{eq}}(\mathbf{x}, t)}{\tau},
\]

where \( f_i^{\text{eq}} \) is the equilibrium distribution function, \( \tau \) is the relaxation time, and \( \mathbf{e}_i \) represents the discrete lattice velocities. The equilibrium distribution function was computed as:

\[
f_i^{\text{eq}} = w_i \rho \left[ 1 + \frac{\mathbf{e}_i \cdot \mathbf{u}}{c_s^2} + \frac{(\mathbf{e}_i \cdot \mathbf{u})^2}{2c_s^4} - \frac{\mathbf{u} \cdot \mathbf{u}}{2c_s^2} \right],
\]

where \( \rho \) is the density, \( \mathbf{u} \) is the velocity, \( w_i \) are the lattice weights, and \( c_s \) is the speed of sound. The boundary conditions were periodic in the \( x \)-direction and fixed velocities at the top and bottom. Shear instability was introduced by perturbing the velocity field with a small amplitude noise to encourage instability growth.

\subsection{Data Generation}

This study focuses on replicating the DiffFluid methodology to evaluate its application on Navier-Stokes and Lattice Boltzmann simulations. We used a consistent setup in both simulations to standardize comparisons and ensure reproducibility. The decision to base our implementation on FNO was based on the suspicion that the DiffFluid paper\cite{luo2024difffluidplaindiffusionmodels} used it as a foundation in their work, although this is not proven. 

\subsubsection{Navier-Stokes}

The Navier-Stokes dataset was generated using a spectral solver on a \(32 \times 32\) grid. Initial vorticity fields were synthesized with random Fourier phases and a prescribed energy spectrum to mimic turbulent-like flows. The forcing function, following the DiffFluid framework, was defined as:

\[
f(x, y) = A\left(\sin(2\pi(x+y)) + \cos(2\pi(x+y))\right)
\]

where \(A\) is the forcing amplitude. Vorticity evolution was computed over 50 timesteps using a Crank-Nicholson integration scheme for time-stepping. This semi-implicit method ensured numerical stability while preserving accuracy for the nonlinear advective and diffusive terms.

After identifying deficiencies in the initial training results, the dataset size was increased from 200 to 300 simulations because the improved training process with early stopping converged quickly enough to allow for the inclusion of more simulations. This adjustment also aimed to bring the dataset size closer to the 1000 simulations used for training in the DiffFluid paper. Each simulation consisted of sequential vorticity snapshots to facilitate temporal learning.

\subsubsection{Lattice-Boltzmann}

LBM was employed to model shear layer instability. A D2Q9 lattice was used on a \(32 \times 32\) grid, with velocity perturbations applied to initialize the instability. The equilibrium distribution function incorporated both velocity and density fields, ensuring consistency with the incompressible Navier-Stokes equations. Boundary conditions at the top and bottom of the domain were enforced to sustain shear flow, while periodicity was maintained along the horizontal axis.

The dataset initially included 200 simulations but was expanded to 300 post-correction. Each simulation comprised density fields captured over 50 timesteps.

\subsection{Model Setup}

The architecture used had a U-Net-inspired backbone and transformer layers for global feature integration. We approximated many aspects of the model in our implementation due to the sparsity of knowledge about the exact methodology used in the Diffluid paper. During training, we experimented with multi-resolution noise to capture fine-scale dynamics. However, we implemented the loss function incorrectly, where predicted noise was directly compared to the ground truth which led to poor optimization, as the target in diffusion models is typically the noise added to the input. After we corrected this issue, we revisited multi-resolution noise but encountered difficulties in parameter tuning, which made it challenging to train the model consistently. Thus, we used methods that align more to standard diffusion model practices that provide more stable results. We opted for standard Gaussian noise and revised the loss function, predicting noise from noisy inputs, which led to better convergence and accuracy.

Various combinations of MSE and L1 losses were tested during training. After experimentation, the final implementation uses MSE and L1 at a \(10\%\) scale. Gradient clipping was employed to mitigate exploding gradients. Additionally, to avoid overfitting, dropout was incorporated within the transformer layers.

The input consists of 10 frames of a 32x32 grid of a feature (vorticity for Navier-Stokes and density for Lattice-Boltzmann), and the goal is to output a single frame after a time step. The input dimensions and number of time steps were reduced to increase computational efficiency. The dataset consisted of 300 samples, which were divided into \(70\%\) training, \(20\%\) validation, and \(10\%\) testing sets. Early stopping was applied once validation loss stabilized around \(0.05\), with most training runs completing within \(50 - 150\) epochs.

\subsection{Model Training}

A weighted combination of mean squared error (MSE) and L1 loss was implemented as the loss function. While MSE focused on achieving global accuracy, L1 loss should help preserve sharp features such as vortices and shear boundaries. To optimize convergence, the AdamW optimizer was employed alongside learning rate scheduling via the ReduceLROnPlateau strategy. Regularization techniques like gradient clipping and dropout, further stabilized the training process under the constraints of limited data. Additionally, early stopping was introduced, halting training approximately between 50 and 150 epochs when the loss stabilized around 0.05.

Initially, the implementation of the loss function was critically flawed: it directly compared predicted noise to the true fluid state \(y\), rather than to the noise embedded in the diffusion process. This oversight led to low training losses (\( \sim 10^{-9}\) in some cases), but poor visual quality in predictions. After correcting this issue, the training pipeline was overhauled. The new pipeline was a lot more robust and had improved accuracy.

\subsection{Evaluation}

The model performance was evaluated using both quantitative metrics and qualitative assessments. The loss function, which is the weighted sum of the mean squared error (MSE) and the L1 loss between predicted and ground-truth states served as a numerical baseline for the quantitative metrics. The MSE error was used to enable precise comparisons of model accuracy across models and with the original paper. Qualitatively, visualizations of flow fields and error maps were employed to examine the model's ability to replicate fine-scale structures such as vortices, instabilities, and sharp transitions.

\subsection{Implementation}

All simulations and model training were conducted in Python. Key libraries included PyTorch for deep learning, NumPy for numerical computations, and Matplotlib for visualization. Computational experiments were performed on NVIDIA GPUs, leveraging CUDA acceleration for efficient tensor operations. The Navier-Stokes solver used fast Fourier transforms for velocity calculations, while the Lattice Boltzmann implementation utilized numpy-based streaming and collision operators.

\end{multicols}
{\color{gray}\hrule}
\begin{center}
\section{Results}
\bigskip
\end{center}
{\color{gray}\hrule}
\begin{multicols}{2}
\subsection{Navier-Stokes}

Initial attempts to replicate the results highlighted in the DiffFluid paper\cite{luo2024difffluidplaindiffusionmodels} failed due to critical errors in the training pipeline. We saw a misalignment of the loss function and the simulation quality, which was caused due to an improper approach in the loss prediction, where the model was trying to predict the ground truth instead of the noise added to the ground truth. While training loss values approached near-zero, the test mean squared error exceeded 50, and visual comparisons of the generated flow fields showed clear artifacts and inaccuracies.

\begin{center}
    \includegraphics[width=\linewidth]{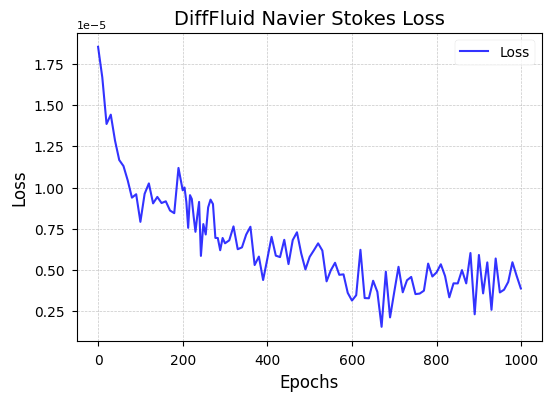}
    \captionof{figure}{Training MSE + L1 loss curve for Navier-Stokes DiffFluid model before the correction}
    \label{fig:navier-stokes-loss-curve___3}
\end{center}

We reconfigured the noise-prediction mechanism by correcting the predicted quantity to noise and adopted a combined loss function of MSE and scaled L1 regularization. Following these corrections, the model exhibited a dramatic improvement in both numerical and visual evaluations. The test loss stabilized around 0.05 after fewer than 50 epochs, and the generated flow fields closely mirrored the ground truth.

\begin{center}
    \includegraphics[width=\linewidth]{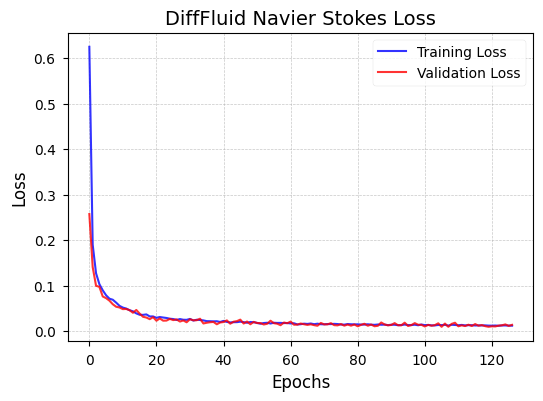}
    \captionof{figure}{Training MSE + L1 loss curve for Navier-Stokes DiffFluid model after the correction}
    \label{fig:navier-stokes-loss-curve-corrected__2}
\end{center}

\subsection{Lattice-Boltzmann}

The second benchmark, based on the Lattice Boltzmann method, initially faced similar challenges. Early training results followed those of the Navier-Stokes simulations, with similar discrepancies between the loss metrics and visual assessments. After correcting the training pipeline similar to the corrections in the Navier-Stokes simulations, the model demonstrated strong adaptability to the lattice-based framework. The corrected training pipeline facilitated sharp and accurate predictions of density and velocity fields across multiple simulation setups. These results validate the versatility of the diffusion transformer across diverse fluid simulation regimes.

\begin{center}
    \includegraphics[width=\linewidth]{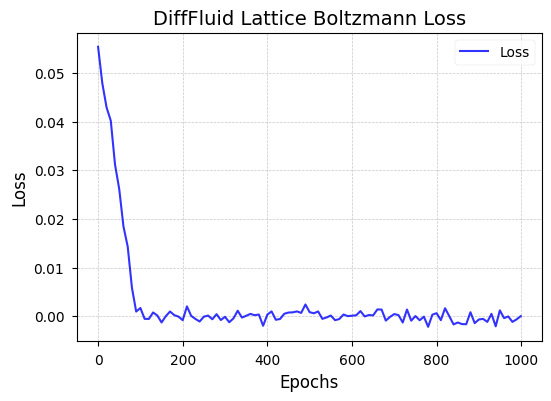}
    \captionof{figure}{Training MSE + L1 loss curve for Lattice-Boltzmann DiffFluid model before the correction}
    \label{fig:navier-stokes-loss-curve__1}
\end{center}

\begin{center}
    \includegraphics[width=\linewidth]{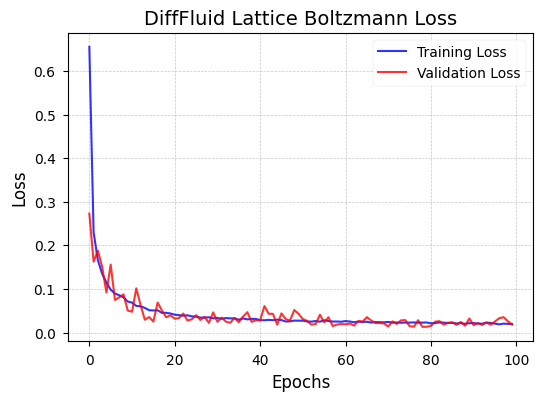}
    \captionof{figure}{Training MSE + L1 loss curve for Lattice-Boltzmann DiffFluid model after the correction}
    \label{fig:navier-stokes-loss-curve-corrected__0}
\end{center}

To make meaningful comparisons to the performance of the DiffFluid model from the paper (with some caveats), we need to calculate the relative L2 error. The L2 error values in table \ref{table:l2} show that the error of around $8\%$ and $6\%$ respectively are relatively close in line with the L2 error values of around $5\%$ reported in the DiffFluid paper.

\begin{center}
\begin{tabular}{c|c}
    DiffFluid model & relative L2 error \\ \hline
    Navier-Stokes & 0.07967 \\
    Lattice Boltzmann & 0.0598
\end{tabular}
\captionof{table}{Relative L2 error for both model variations}
\label{table:l2}
\end{center}

\subsection{Visualization}

Visual comparisons of the ground truth and predictions, and error maps further highlight the improvements introduced by the corrected pipeline. Error maps reveal a significant reduction in high-error regions and a faster convergence compared to the results obtained from the incorrect training pipeline. Flow fields generated by the diffusion transformer align closely with ground truth data, maintaining sharpness and coherence in regions of high-gradient flow features.

\begin{center}
    \includegraphics[width=\linewidth]{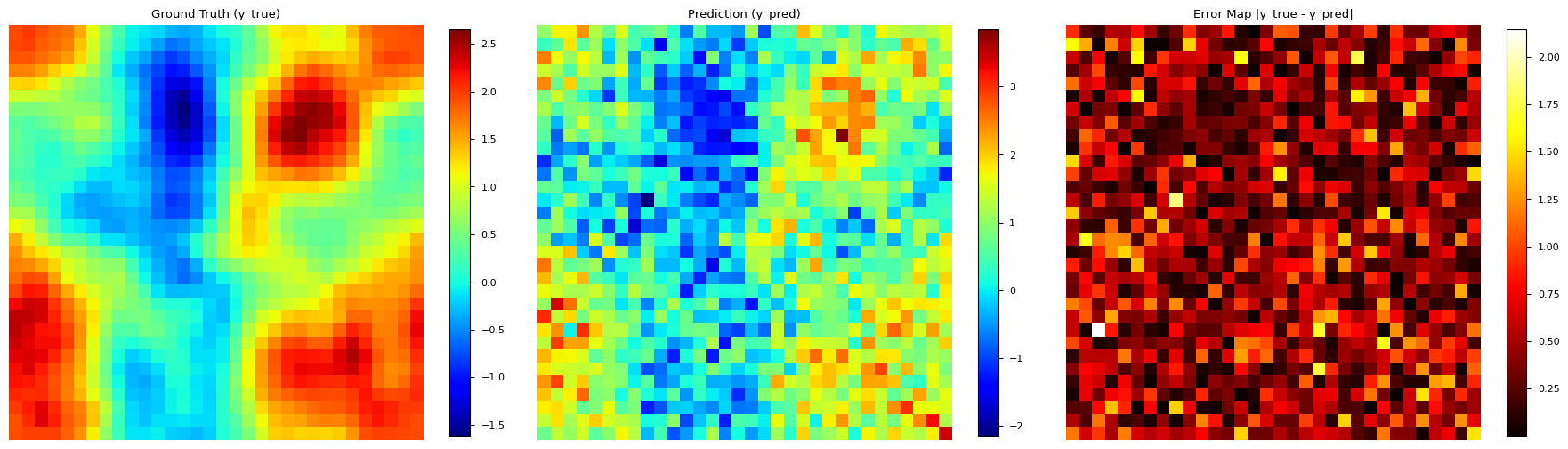}
    \captionof{figure}{Ground Truth vs. Prediction from the Navier-Stokes model on a randomly selected sequence from the test data after training for 1000 epochs on the wrong method}
    \label{fig:navier-stokes-comparison_1}
\end{center}

\begin{center}
    \includegraphics[width=\linewidth]{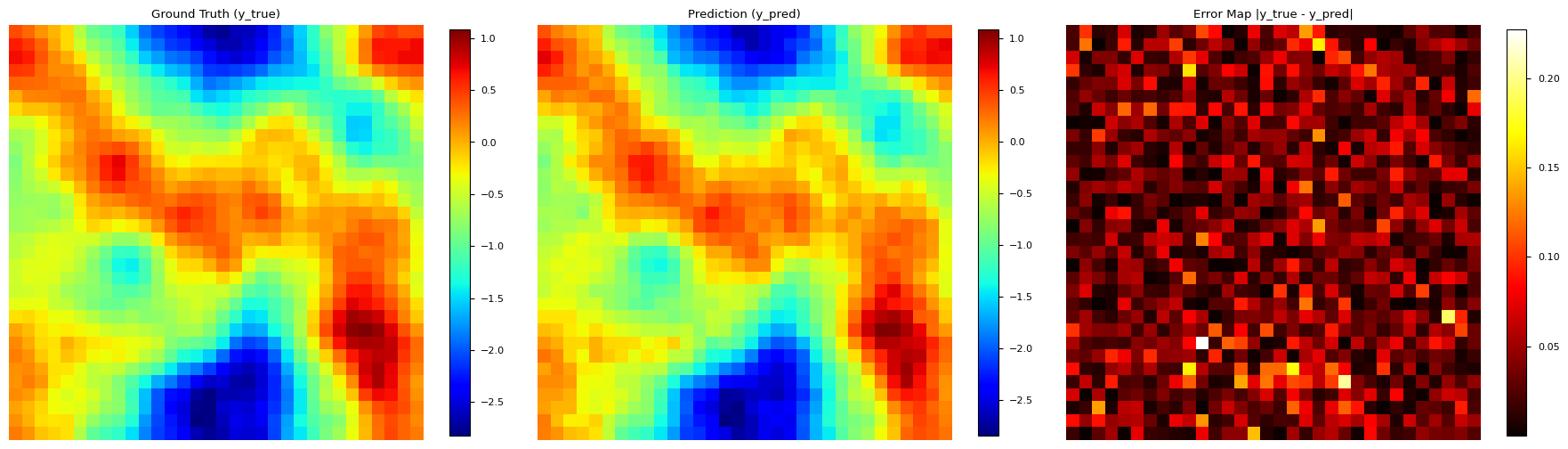}
    \includegraphics[width=\linewidth]{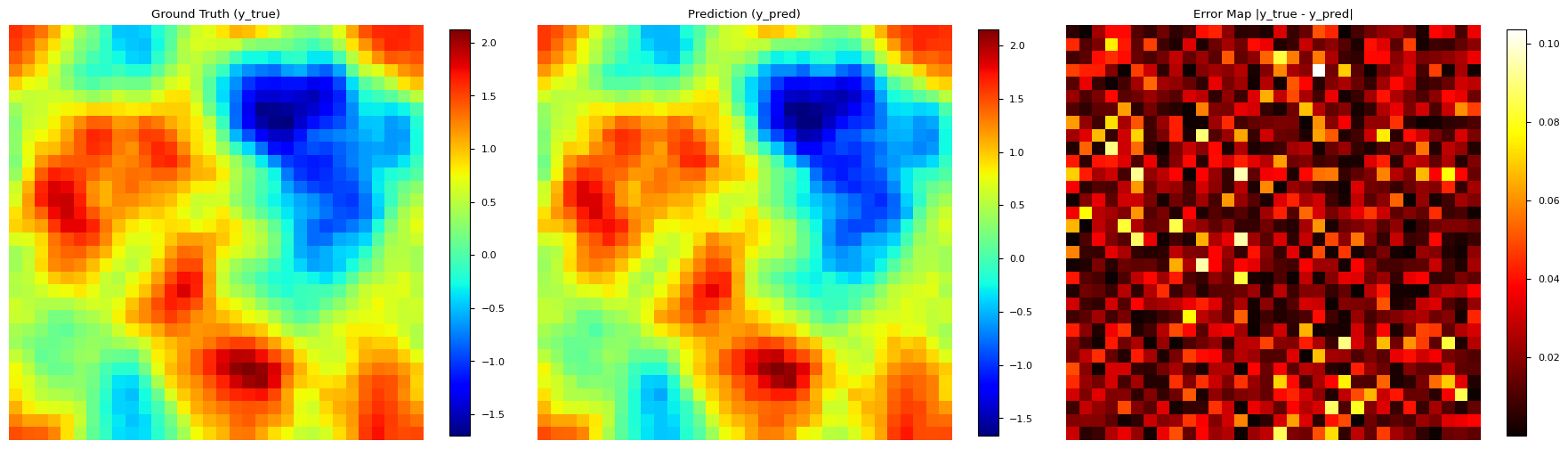}
    \includegraphics[width=\linewidth]{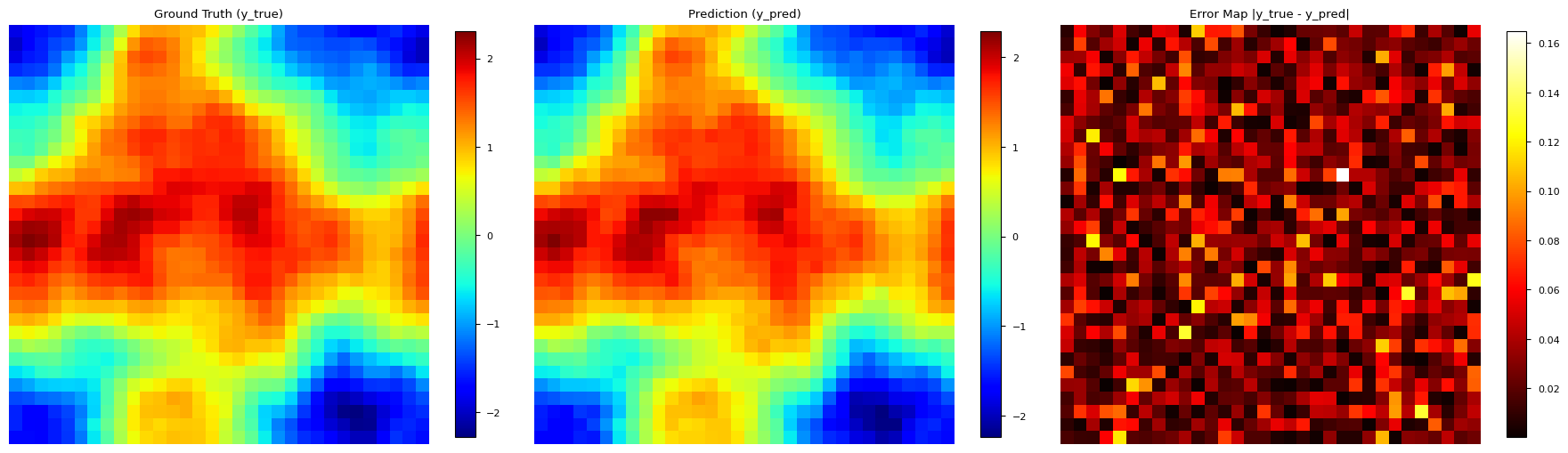}
    \captionof{figure}{Ground Truth vs. Prediction from the Navier-Stokes model on a randomly selected sequence from the test data after training for 200 epochs with the corrected method}
    \label{fig:navier-stokes-comparison_0}
\end{center}

\begin{center}
    \includegraphics[width=\linewidth]{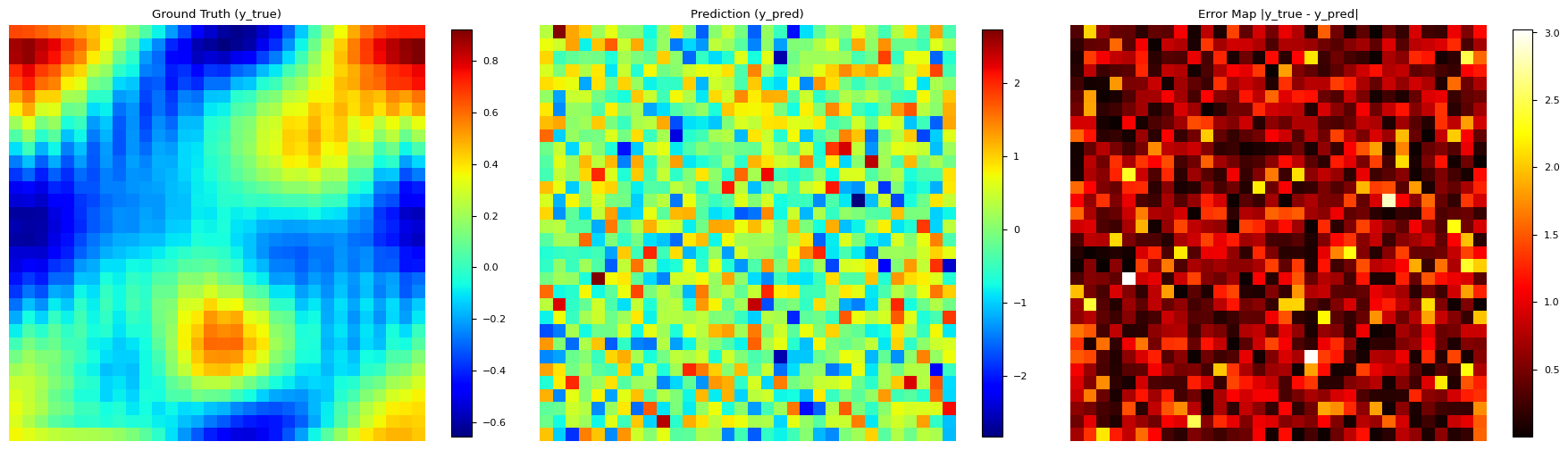}
    \captionof{figure}{Ground Truth vs. Prediction from the Lattice-Boltzmann model on a randomly selected sequence from the test data after training for 1000 epochs on the wrong method}
    \label{fig:navier-stokes-comparison}
\end{center}

\begin{center}
    \includegraphics[width=\linewidth]{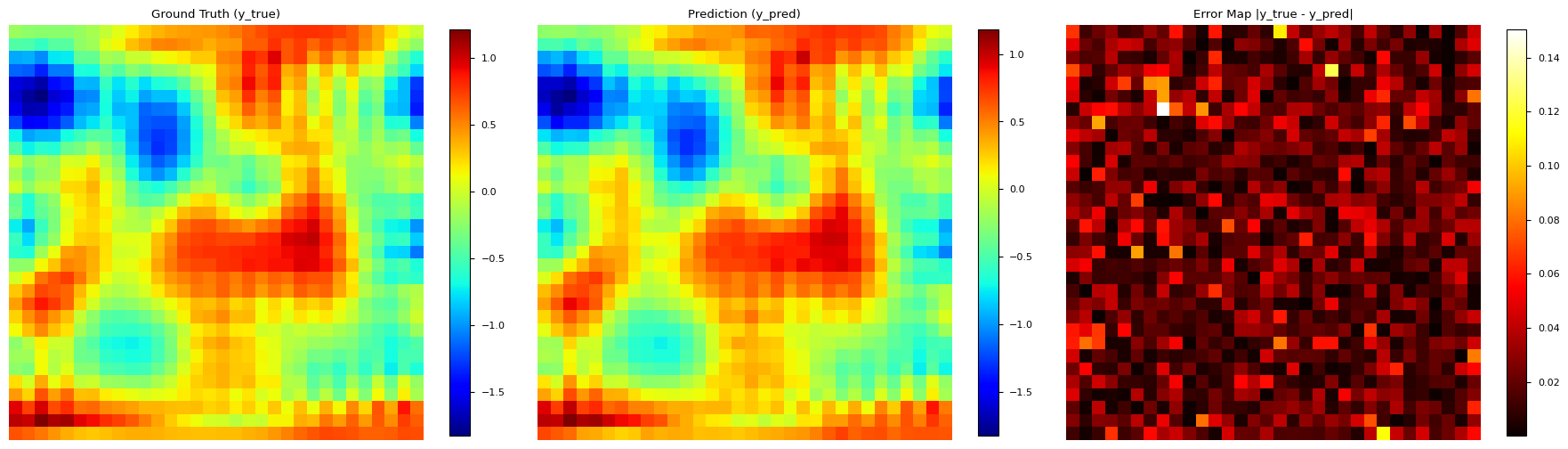}
    \includegraphics[width=\linewidth]{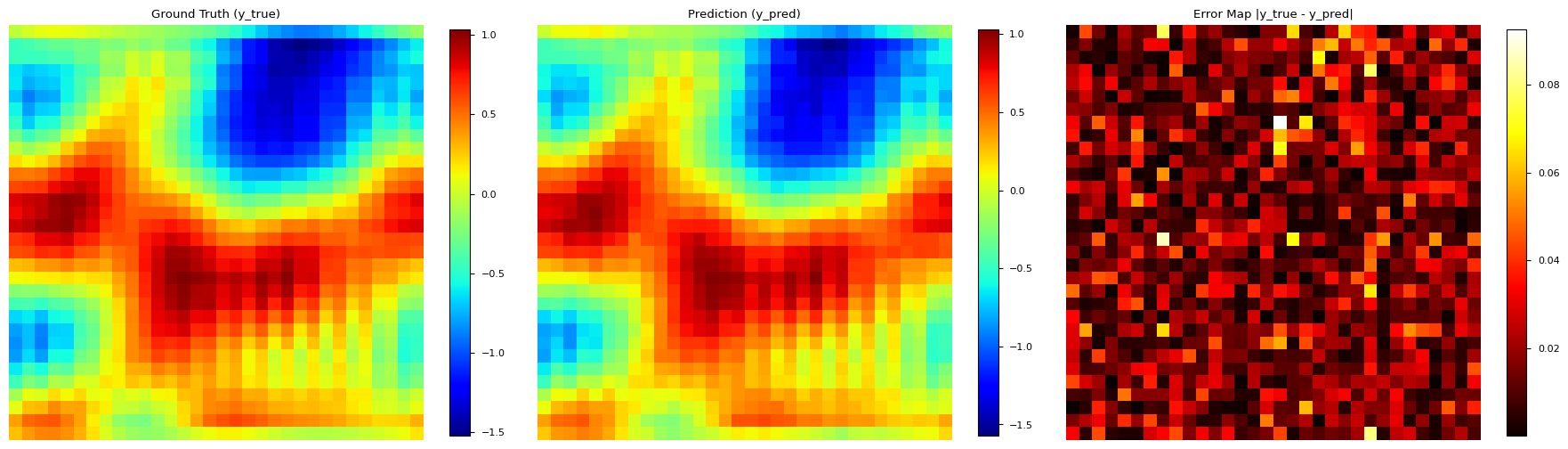}
    \includegraphics[width=\linewidth]{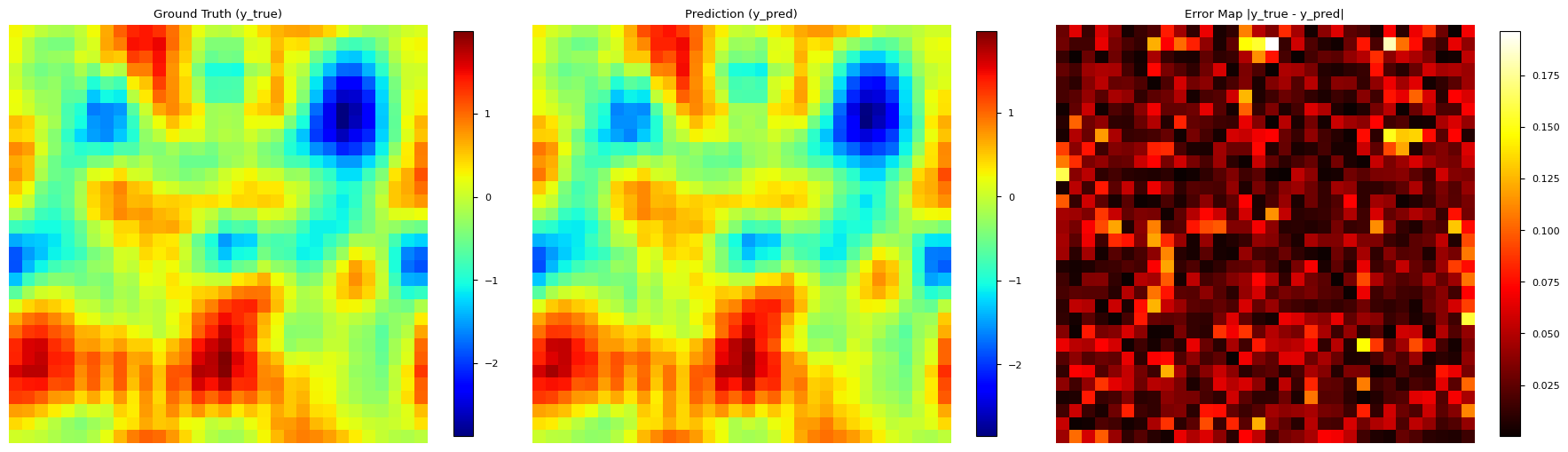}
    \captionof{figure}{Ground Truth vs. Prediction from the Lattice-Boltzmann model on a randomly selected sequence from the test data after training for 200 epochs with the corrected method}
    \label{fig:navier-stokes-comparison_2}
\end{center}



\end{multicols}
{\color{gray}\hrule}
\begin{center}
\section{Discussion}
\bigskip
\end{center}
{\color{gray}\hrule}
\begin{multicols}{2}

The results presented in this study highlight the adaptability and potential of diffusion-based transformer architectures for modeling fluid dynamics, specifically in the contexts of Navier-Stokes and Lattice Boltzmann simulations. We systematically addressed challenges faced during the training of the models and were able to achieve a reliable replication of some of the outcomes described in the DiffFluid framework\cite{luo2024difffluidplaindiffusionmodels} given the computational limitations and extended its application to new simulation types.

We discovered a fundamental error in noise prediction within our initial implementation only after a series of long training runs of 1000 epochs. Before fixing the issue, the training loss decreased to very low values, while the visual output of the generated flow fields was at times very far off from what we expected. Since this was not a one-to-one replication of the paper and we had made some design decisions to reduce the computational costs, the problem could not be localized straightway. Once the issue was solved, the updated training strategy, which combined MSE with a scaled L1 loss, significantly improved the model, to the point where it was able to generate realistic next frames in the simulation. This highlights how vital it is to verify each aspect of the model before starting a long training run which costs time and computational resources.

In the case of Navier-Stokes simulations, the corrected pipeline achieved a marked improvement in both qualitative and quantitative performance metrics. The visual inspection of flow fields revealed the model's capability to capture key features of turbulent behavior with decent quality given its relative simplicity. The stabilized training loss at approximately 0.05 and the associated error maps indicated that the model could reliably generalize across test scenarios despite the relatively modest dataset size and computational budget. Our results demonstrate the robustness of the diffusion transformer architecture and its ability to incorporate complex interactions within fluid dynamics.

The application of the model to Lattice Boltzmann simulations further validated its versatility. We adapted the same architectural framework with minimal adjustments to model interactions distinctive of lattice-based approaches. The ability to maintain performance consistency across simulation types highlights the generalizability of the diffusion-based transformer design, which promises potential for broad applications in computational fluid dynamics.

While we were at least partially successful in our goal of showing the adaptability of the DiffFluid model, our project was not without its limitations. The initial error in the training pipeline and the resulting troubleshooting were significant setbacks that cost us time. Training diffusion models limited the scope of our experimentation since they usually take longer to train and are more sensitive to the specific setup and parameters. Larger datasets and more extensive training could potentially increase performance with fine-grained flow details. While early stopping mitigated overfitting risks, it also restricted the exploration of long-term temporal dependencies within the simulated systems. Given the generalizability of the architecture observed, a possible extension to this work could focus on leveraging transfer learning and adaptive learning techniques to overcome these barriers.

Another potential issue is that the model might be learning to copy the last given frame in the 10-frame sequence since this is very close to the predicted frame in most circumstances. Our scope was too limited to predict further than the next frame due to the training requirements. Because these types of time-based simulations usually don't have drastic changes between individual frames, unless \(dt\) is very high which would in turn also make the simulation less realistic, there is a chance that the model recognizes that outputting the last frame is close enough to the target. This would mean the model doesn't learn the underlying physical relationships. 

But overall, our findings are in line with the DiffFluid paper. With more extensive training and further adjustments to the parameters and implementation of our replicated DiffFluid model, we would hopefully find similar results. This is supporting evidence to show that further research into diffusion based models for simulating physical phenomena is warranted. A clear path forward would be to add simulation parameters into the model so it could be used trained on a variety of parameter combinations to get a more generalized model.

\end{multicols}
{\color{gray}\hrule}
\begin{center}
\section{Conclusions}
\bigskip
\end{center}
{\color{gray}\hrule}
\vspace{0.5cm}
This study demonstrates that diffusion-based transformers are effective and adaptable tools for modeling fluid dynamics. Our findings showcase the model's flexibility across different simulation paradigms and training pipelines and provide further evidence of the potential of using ML architectures like transformers to solve problems in Computational Physics. There is likely a great potential for the refinement and expansion of these methods to solve problems in broad domains of science and engineering that have eluded a solution by traditional methods.
\bibliographystyle{unsrt}
\bibliography{references}
\end{document}